\documentclass[aps,prl,reprint,superscriptaddress,citeautoscript]{revtex4-2}
\usepackage[version=4]{mhchem}
\usepackage{amsmath, amssymb}  
\usepackage{graphicx}    
\usepackage{lipsum}
\usepackage{color}  
\usepackage[dvipsnames]{xcolor}

\begin{document}
\title{Hidden Gauge Freedom in Complex-Pole Hierarchical Equations of Motion}
\author{Tianchu Li}
\affiliation{Department of Chemistry, University of Colorado Boulder, Boulder, Colorado 80309, USA\looseness=-1}

\author{Andr\'es Montoya-Castillo}
\email{Andres.MontoyaCastillo@colorado.edu}
\affiliation{Department of Chemistry, University of Colorado Boulder, Boulder, Colorado 80309, USA\looseness=-1}
\date{\today}

\begin{abstract}
While complex-pole hierarchical equations of motion (HEOM) have dramatically expanded the reach of numerically exact quantum dynamics simulations of open quantum systems, they suffer from numerical instabilities rooted in the non-Hermitian structure of their Liouvillian. Yet, the origin of this structure remains obscure. Here, we report a previously unknown gauge freedom in complex-pole HEOM: a continuous family of analytically equivalent Liouvillians, all encoding the same bath correlation function, whose numerical properties vary dramatically. This gauge controls both the eigenspectrum and non-normality of the hierarchy generator, revealing spectral divergence and non-normal error amplification as two distinct instability mechanisms. By optimizing this gauge, we introduce GO--HEOM, which eliminates divergences in strongly coupled Brownian oscillator environments and extends numerically exact simulations of sub-Ohmic dynamics---including through the delocalized-to-localized quantum phase transition---to previously inaccessible coupling strengths. Because this gauge transformation is independent of the bath-correlation decomposition scheme, our GO--HEOM becomes a general, broadly compatible strategy for accessing numerically exact quantum dynamics of open quantum systems over arbitrary coupling and highly non-Markovian regimes.
\end{abstract}

\maketitle

\textit{Introduction} --- Accurate simulation of open quantum system dynamics is essential for understanding energy transfer in photosynthetic systems~\cite{tanaka09, tanaka10, liu21, huang2024simulation, li2026molecular, strathearn18, leng18}, charge transport in solids and molecular junctions~\cite{hartle13, song15, yan19, xing22, jankovic2022spectral, kloss19, li24, jankovic2025charge, bhattacharyya2025nonequilibrium, li21}, and decoherence in quantum devices~\cite{zhang24, nakamura2025impact, nakamura2026entanglement, chen2026simulation, li2026tensor}. The hierarchical equations of motion (HEOM) have emerged as a powerful and versatile tool that provides numerically exact, non-perturbative solutions to these problems~\cite{tanimura89, tanimura06, yan04, tanimura20, bai24}. Despite their success, HEOM simulations remain challenging due to the rapid growth of the hierarchy and limitations to specific forms of the spectral density. Over the past decade, on-the-fly filtering~\cite{shi09b}, tensor-network formulations~\cite{shi18,borrelli21,yan21,ke22,ke2023tree,chen2025tree,guan24}, and alternative bath-correlation decompositions~\cite{chen22,zhang2025minimal,xu22,hunt2026exploiting} have dramatically improved efficiency, with the latter also having broadened the applicability of HEOM to problems with arbitrary spectral densities. However, this breadth comes at a cost: stability. These numerical instabilities are now the major limitation preventing access to some of the most physically interesting regimes of open quantum dynamics~\cite{li22,yan21a,dunn19,firmino16}. The sub-Ohmic spin-boson model, for instance, hosts a quantum phase transition between delocalized and localized behavior whose precise dynamical signatures at strong coupling have remained beyond HEOM's reach~\cite{winter2009quantum,goulko2025transient,wang10a,bulla03}. Similarly, strongly coupled Brownian oscillator environments---directly relevant to electron transfer in solution---produce divergent HEOM dynamics at physically realistic reorganization energies.

These alternative bath correlation decompositions rely on complex poles~\cite{chen22,zhang2025minimal,nakatsukasa2018aaa,xu22,dan23a} that yield highly non-Hermitian HEOM Liouvillians whose numerical instability arises from eigenvalues with positive real parts that lead to divergent dynamics~\cite{yan21,li22}. We show that these HEOM reformulations exhibit significant non-normality that can accelerate error growth over intermediate-to-long times (see Fig.~\ref{fig3}). However, the connection between the structure of the complex-pole HEOM and these numerical properties remains poorly understood. This lack of understanding poses a major challenge for the stable simulation of complex-pole HEOM in strongly coupled, long-memory, and highly structured environments.

\begin{figure}[b]
\centering
\includegraphics[width=8.70cm]{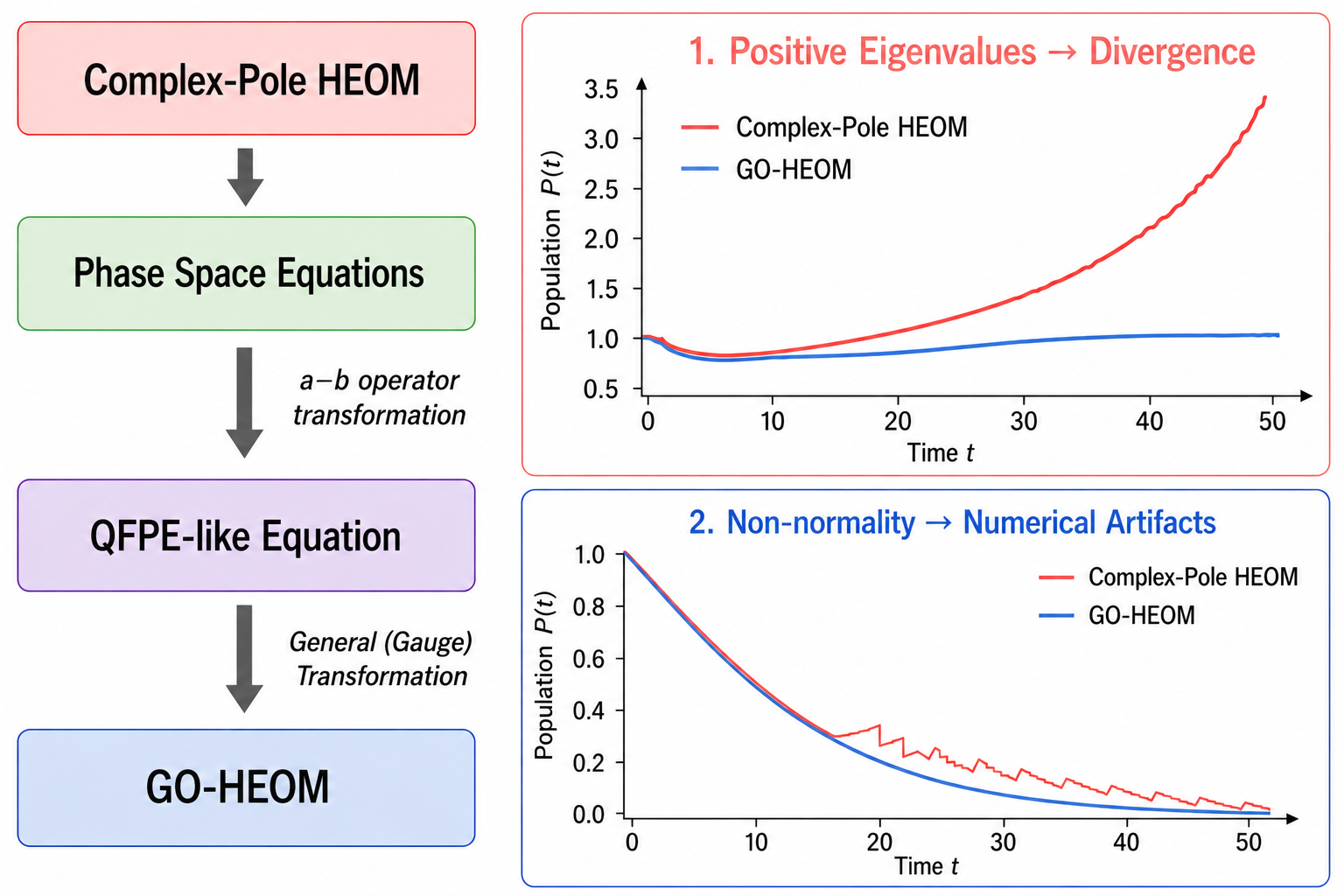}
\vspace{-20pt}
\caption{
Schematic derivation of the GO--HEOM and the two major instability mechanisms in complex-pole HEOM.
}
\vspace{-10pt}
\label{fig0}
\end{figure}

In this Letter, we show that complex-pole HEOM possesses a previously unknown phase or gauge freedom~\footnote{We use the term gauge in the sense standard in physics: a transformation acting on unphysical (here, auxiliary) degrees of freedom that leaves all physical observables invariant while reshaping the mathematical representation of the generator. This is structurally analogous to the Lindblad jump-operator gauge~\cite{reiter2012effective}, and distinct from the trivial rescaling of ADOs discussed in Ref.~\cite{dunn19}.} that preserves the bath correlation function while continuously reshaping the structure of the HEOM Liouvillian. As such, HEOM exhibits gauge invariance, but this gauge tunes its numerical stability. We show that this freedom controls the spectral properties and non-normality of the resulting HEOM, and uncover the connections between HEOM Liouvillian structure, non-normality, and numerical stability. We establish that one can continuously improve the numerical stability of complex-pole HEOM by tuning the gauge, yielding what we call gauge-optimized (GO)-HEOM (see Fig.~\ref{fig0}), which enables stable simulations in strongly coupled and highly non-Markovian environments. We illustrate these advantages on the Brownian oscillator and sub-Ohmic spectral densities, where our GO-HEOM tames previously fatal numerical instabilities, broadening HEOM's applicability to formerly inaccessible chemical and physical systems. More broadly, these results reveal that the structure of the HEOM Liouvillian is not determined by physics alone---it carries a representational degree of freedom that has been invisible in all prior formulations. Exploiting this freedom not only cures known instabilities but suggests that analogous hidden freedoms may exist in other exact methods for open quantum dynamics.

\textit{Model system and method} --- To expose the structural origin of the gauge freedom in complex-pole HEOM, we first reformulate the hierarchy in a phase-space representation, where the auxiliary density operators (ADOs) are mapped onto auxiliary phase-space variables that may be associated with fictitious bosonic modes. A linear transformation of these auxiliary bosons yields a Fokker–Planck-like equation of motion, which reveals the dissipative and oscillatory components of the hierarchy generator. Within this phase-space framework, we uncover a continuous gauge freedom parameterized by $\phi$ that continuously reshapes the structure of the HEOM Liouvillian. Mapping the resulting phase-space equation back to the hierarchy representation then yields a family of $\phi$-dependent complex-pole HEOMs. We then show that one can maximize the numerical stability of HEOM by optimizing $\phi$, yielding our GO-HEOM.

We illustrate the resulting framework using the canonical spin--boson model---a minimal model of quantum relaxation that describes a two-level system linearly coupled to a Gaussian bath~\cite{leggett87, may04},
\begin{align}
H = \frac{\epsilon}{2}\sigma_z + V\sigma_x
+\frac{1}{2}\sum_j \left[
p_j^2
+\omega_j^2
\left(x_j-\frac{c_j\sigma_z}{\omega_j^2}\right)^2
\right].
\end{align}
Here, $\epsilon$ and $V$ denote the energy bias and diabatic coupling of the two-level system, respectively, while $\sigma_{x,z}$ are Pauli matrices. In all our simulations, we set $\epsilon = 1.0$ and $V = 1.0$. $x_j$, $p_j$, and $\omega_j$ represent the coordinate, momentum, and frequency of the $j$th bath oscillator, respectively. The spectral density, $J(\omega) = \frac{\pi}{2} \sum_j \frac{c_j^2}{ \omega_j} \delta(\omega - \omega_j)$, fully characterizes the bath and its coupling to the system, and determines the bath correlation function,
\begin{align}
    C(t) = \langle F(t)F(0) \rangle    
    = \frac{1}{\pi} \int_0^\infty d\omega \,
    \frac{J(\omega)e^{-i\omega t}}{1-e^{-\beta\omega}},
\label{eq:bcf}
\end{align}
where $F=\sum_j c_j x_j$ is the collective bath coupling and $\beta=1/(k_B T)$ is the inverse thermal energy.

In complex-pole HEOM~\cite{liu14,tanaka09,xu22}, one decomposes the bath correlation function into a sum of exponentials,
\begin{align}
C(t)=\sum_k d_k e^{-z_k t},
\end{align}
and obtains the complex coefficients $d_k$ and $z_k$ using the frequency-domain AAA decomposition~\cite{xu22} or the time-domain Prony fitting method~\cite{chen22}. The explicit form of the resulting complex-pole HEOM is given in Eq.~(S6) of the Supplemental Material (SM).

To uncover the gauge freedom hidden in complex-pole HEOM,  we reformulate the hierarchy in a phase-space representation~\cite{shi09c,liu14,yan20,li22,li25,xu2023universal}. Earlier works leveraged this phase-space reformulation of HEOM to elucidate its connection to the quantum Fokker-Planck equation (QFPE)\cite{garg85,leggett87} and to derive alternative HEOM formulations for Lorentzian spectral densities\cite{li22,ikeda20}. While earlier work exploited a Fokker–Planck reformulation specifically for the Drude–Lorentz spectral density to derive a numerically stabilized HEOM~\cite{shi09c}, we show that this phase-space perspective, generalized to arbitrary complex-pole decompositions, reveals a continuous family of valid HEOM formulations---a gauge freedom that was invisible at the level of any single decomposition and that previous stabilization strategies~\cite{shi09c, dunn19} accessed only implicitly and partially. Our reformulation reveals that there is an infinite family of rotations of the Fokker-Planck operator that, while analytically equivalent, have different numerical properties. 

We illustrate our argument by first considering a single exponential contribution to the bath correlation function, $C(t) = d_0 e^{-z_0t}$, introducing two canonical ladder operators $c_1$ and $ c_2$ with $[c_i, c_j^+] =\delta_{i,j}$, and expanding the auxiliary phase-space distribution $\rho_s(x,p;t) = \sum_{m,n} (m!n!)^{-1/2} \rho_{m,n}(t)\phi_{m,n}(x,p)$ in the Gaussian basis $\phi_{m,n}(x,p)= (2\pi)^{-1/2}e^{-S}({c}_1^+)^m({c}_2^+)^n |0,0\rangle$ with $S = (x^2 +p^2)/4$ and $|0,0\rangle = (1/\sqrt{2\pi})e^{-S}$. Under this mapping, ADO indices $(m,n)$ correspond to the occupation numbers of the ladder operator modes. Thus, the hierarchy truncation depth becomes equivalent to a Fock-space cutoff of the auxiliary phase-space representation. 

The complex-pole HEOM thus obeys the following phase-space equation of motion:
\begin{align}
    \dot{\rho}_s(x,p)
    =&\;-e^{-S}
        \Bigl(i\mathcal{L}
             + z_0{c}_{1}^+{c}_{1}
             + z_0^*{c}_{2}^+{c}_{2}\Bigr)
        e^{S}
        \rho_s \nonumber\\
    & -ie^{-S}
       \bigl({c}_{1}+{c}_{2}\bigr)
       e^{S}
       \bigl[\sigma_z,\rho_s\bigr]
    \nonumber\\
    &-ie^{-S}
       \Bigl[
         d_0\,{c}_{1}^+
         e^{S}
         \sigma_z\,\rho_s -d_0^*{c}_{2}^+
         e^{S}
         \rho_s\,\sigma_z
       \Bigr].
    \label{eq:pseom}
\end{align}
We refer the reader to the SM for details on this phase-space construction. By transforming the bosons
\begin{subequations} \label{eq:rotation-zero}
\begin{align}\label{eq:rotation0}
     c_{1}^+ &=-\frac{\gamma-i\omega}{\Omega}a^++ b^+, & c_{2}^+ &=-\frac{\gamma+i\omega}{\Omega} a^++ b^+, 
    \\
     c_{1}&=\frac{\Omega}{2i\omega}a+\frac{\gamma+i\omega}{2i\omega}b, & c_{2}&=-\frac{\Omega}{2i\omega} a-\frac{\gamma-i\omega}{2i\omega} b,
    \label{eq:rotation1}
\end{align}
\end{subequations}
where $\gamma = {\rm Re}\ z$, $\omega = {\rm Im}\ z$, $\Omega = \sqrt{\gamma^2+\omega^2}$ and $(a^+,a),(b^+,b) $ are bosonic creation and annihilation operators (see SM for the phase-space definitions of $a$ and $b$), the complex-pole contribution term becomes 
\begin{align}
    z_0 c^+_1 c_1 + z_0^*   c_2^+  c_2
    \mapsto
    2\gamma  b^+  b
    +\Omega( a  b^+ -  a^+ b) = \mathcal{L}_{\rm FP}.
\end{align}
This term is equivalent to the Fokker-Planck Liouvillian ($\mathcal{L}_{\rm FP}$) in the QFPE~\cite{garg85,leggett87} (see SM for the explicit form of this QFPE-like equation). 

Importantly, while the transformation in Eq.~(\ref{eq:rotation-zero}) constitutes one map of the $(a,b)$-bosons to an instantiation of HEOM, many linear transformations create equally valid instantiations of HEOM. This nonuniqueness of the hierarchy is analogous to the equivalence of various basis expansions for the same function or resummations of moments for the same functional average. In short, $\mathcal L_{\rm FP}$ possesses a continuous family of equivalent ladder-operator representations that generate identical phase-space equations of motion. We parameterize this hidden freedom by an angle $\phi$. Motivated by Eq.~(\ref{eq:rotation-zero}), we therefore propose the more general transformation
\begin{align}
    \label{eq:gr}
    &\begin{pmatrix}
    \tilde c_1^+\\
    \tilde c_2^+
    \end{pmatrix} = {\bf M}\begin{pmatrix}
    a^+\\
    b^+
    \end{pmatrix}, \quad
    \begin{pmatrix}
    \tilde c_1\\
    \tilde c_2
    \end{pmatrix} = ({\bf M}^{-1})^T\begin{pmatrix}
    a\\
    b
    \end{pmatrix},
\end{align}
with
\begin{equation}
\label{eq:gr1}
    {\bf M} = \begin{pmatrix}
    -e^{-i\phi} & 1\\
    -e^{i\phi} & 1
\end{pmatrix},
\end{equation}
where ${\bf M}$ is invertible for $\phi \neq n\pi$, so that the inverse transformation is well defined. This allows us to rewrite the Fokker--Planck operator as
\begin{align}
    2\gamma b^+b + \Omega(ab^+-a^+b)
    &=
    A_{11}\tilde c_1^+ \tilde c_1
    +A_{22}\tilde c_2^+ \tilde c_2\nonumber \\
    &+A_{12}\tilde c_1^+ \tilde c_2
    +A_{21}\tilde c_2^+ \tilde c_1 ,
\end{align}
where the coefficients are defined in the SM. By reversing the derivation leading to Eq.~(\ref{eq:pseom}), redefining the basis as $\phi_{m,n}(x,p)=\frac{1}{\sqrt{2\pi}}e^{-S}(\tilde c_1^+)^m(\tilde c_2^+)^n|0,0\rangle$,
and, generalizing the construction to the multi-exponential case, we obtain the $\phi$-dependent complex-pole HEOM:
\begin{widetext}
    \begin{align}
    \label{eq:gauge-dependent-HEOM}
    \dot{\rho}_{\mathbf m, \mathbf n} =& -\left(i\mathcal L + \sum_k A_{11,k} m_k +\sum_kA_{22,k}n_k \right)\rho_{\mathbf m, \mathbf n} -\sum_k \left(A_{12,k}m_k\rho_{{\bf m}_k^-,{\bf n}_k^+}+A_{21,k}n_k\rho_{{\bf m}_k^+,{\bf n}_k^-}\right)\nonumber \\
    &-\sum_k d_k \frac{e^{i\phi_k}-e^{-i\theta_k}}{2\sin\phi_k}m_k\sigma_z \rho_{{\bf m}_k^-,\bf n} 
    + \sum_k   d_k^*\frac{e^{i\phi_k}-e^{i\theta_k}}{2\sin\phi_k}m_k\rho_{{\bf m}_k^-,\bf n} \sigma_z\nonumber -i\sum_k \left[\sigma_z, \rho_{{\bf m}_k^+,\bf n}\right]\\  
    &-\sum_k d_k \frac{e^{-i\theta_k}-e^{-i\phi_k}}{2\sin\phi_k}n_k\sigma_z \rho_{{\bf m},{\bf n}_k^-}
    + \sum_k   d_k^*\frac{e^{i\theta_k}-e^{-i\phi_k}}{2\sin\phi_k}n_k\rho_{{\bf m},{\bf n}_k^-} \sigma_z-i\sum_k \left[\sigma_z, \rho_{{\bf m},{\bf n}_k^+}\right].
\end{align}
\end{widetext}
Here, $\mathbf m={m_1,m_2,\dots}$ and $\mathbf n={n_1,n_2,\dots}$ are integer indices labeling the ADOs, $\theta_k=\arctan(\omega_k/\gamma_k)$, and $k$ enumerates the members of the complex-pole decomposition of an arbitrary spectral density: $C(t) = \sum_k d_k e^{-z_k t}$. Equation (\ref{eq:gauge-dependent-HEOM}) represents the central result in our work and embodies the infinite family of HEOM variants parameterized by the gauge freedom $\boldsymbol{\phi} \equiv (\phi_1, \dots, \phi_k, \dots )$.

One can recover the conventional complex-pole HEOM as a special case of our gauge-HEOM by setting $\phi_k=\theta_k$. Different values of $\phi_k$ generate analytically equivalent versions of HEOM for the same bath correlation function and yield identical dynamics in the converged limit. In addition, the gauge-HEOM equations are invariant to independent sign changes of the gauge parameters, $\phi_k\rightarrow-\phi_k$. Thus, $\boldsymbol{\phi}$ is not an additional physical parameter, but a representational degree of freedom that can be used to reshape the HEOM Liouvillian and, as we show below, its numerical stability. Importantly, under the generalized transformation of Eqs.~(\ref{eq:gr}), the coefficients of the raising terms, $\rho_{{\bf m}_k^+,\mathbf n}$ and $\rho_{{\bf m},\mathbf n_k^+}$, remain unity. This property ensures that the resulting gauge-HEOM can be derived from a corresponding generalized decomposition of the bath correlation function (see SM).

\begin{figure}[ht!]
    \centering
    \includegraphics[width=8.70cm]{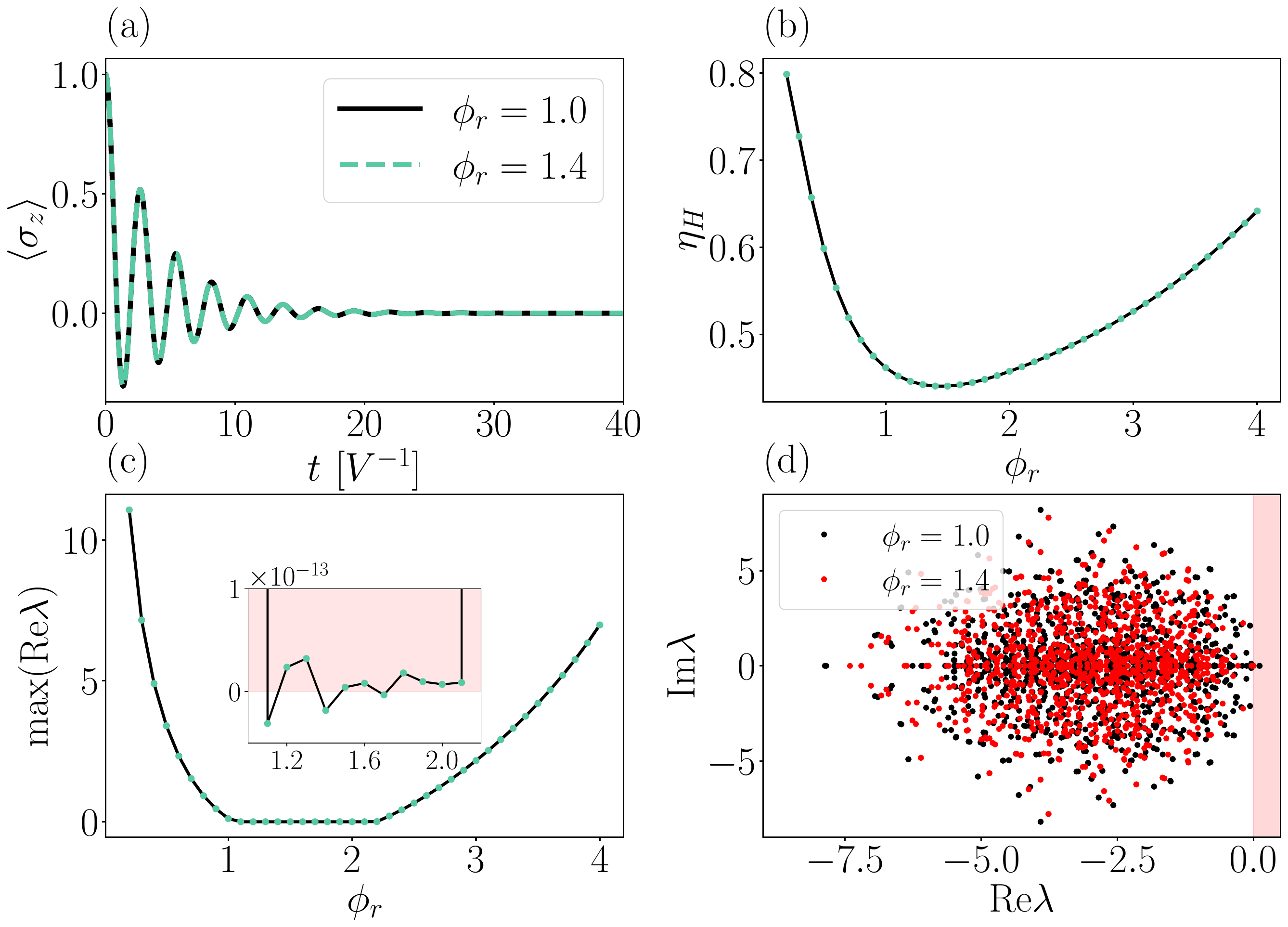}
    \caption{(a) $\langle\sigma_z(t)\rangle$ under an intermediate coupled BO spectral density ($p_0=1.0$) for conventional complex pole HEOM ($\phi_r =1.0$) and GO--HEOM ($\phi_r =1.4$). In the strong-coupling regime of the BO spectral density ($p_0 = 3.0$). 
    (b) Henrici departure from normality as a function of the gauge-scaling parameter $\phi_r$. 
    (c) Maximum real part of the eigenvalue spectrum as a function of $\phi_r$. 
    (d) Eigenspectra for $\phi_r = 1.0$ and $\phi_r = 1.4$.
    }
    \label{fig1}
\end{figure}

\textit{Results} --- To arrive at our GO-HEOM and simulate a system with it, we make our mode-specific gauge transformation proportional to this phase,
$\phi_k = \phi_r \theta_k$, which facilitates gauge-transforming all modes simultaneously with one global scaling parameter, $\phi_r$, over which we can optimize the numerical properties of the resulting gauge-HEOM. Setting $\phi_r = 1.0$ recovers conventional HEOM. Identifying the optimal $\phi_r$ requires scanning a one-dimensional parameter space and, for each candidate value, computing the eigenspectrum or a short test trajectory of the resulting HEOM---a cost that is negligible compared to a full propagation. 

We begin by demonstrating that our gauge-HEOM is numerically exact. To achieve this, we consider a spin--boson model coupled to a Brownian oscillator (BO) spectral density, which serves as a standard benchmark for non-Markovian dissipative dynamics~\cite{ito16,meier99}, 
\begin{align}
J(\omega) = \frac{p_0\omega}{[(\omega+\omega_0)^2+\Gamma^2][(\omega-\omega_0)^2+\Gamma^2]},
\end{align}
where the reorganization energy is defined as $\lambda = p_0/(\Gamma\sqrt{\Gamma^2+\omega_0^2})$. We fix the BO frequency and damping parameter to $\omega_0 = 0.3$ and $\Gamma = 0.5$, corresponding to a low-frequency solvent environment relevant to electron-transfer processes~\cite{firmino16}. Figure~\ref{fig1}(a) compares the dynamics of $\langle\sigma_z(t)\rangle$ for $\phi_r = 1.0$ and $\phi_r = 1.4$ in the intermediate-coupling regime ($p_0 = 1$). The indistinguishable dynamics confirm that the gauge transformation preserves the exact quantum dynamics.

We then interrogate how the gauge transformation modifies the eigenspectrum and non-normality of the HEOM Liouvillian, and how these changes influence numerical stability. In Fig.~\ref{fig1}(b), we quantify the non-normality of the HEOM generator using the Henrici departure from normality~\cite{trefethen1993hydrodynamic,schmid2007nonmodal}, $\eta_H =\sqrt{\|A\|_F^2-\sum_i |\lambda_i|^2}/\|A\|_F$, where the HEOM with truncation can be written in the form $\dot{ \boldsymbol{\rho}}_{\mathbf n} = \bf A \boldsymbol{\rho}_{\mathbf n}$, $\|\cdot\|_F$ denotes the Frobenius norm, and $\lambda_i$ are the eigenvalues of $\bf A$. The Henrici departure from normality provides a quantitative measure of the non-normality of the HEOM Liouvillian. Larger values of $\eta_H$ indicate stronger non-normality, which can lead to enhanced transient amplification of numerical errors during propagation. Figure~\ref{fig1}(b) shows that $\eta_H$ depends sensitively on the gauge-scaling parameter $\phi_r$ and reaches a minimum near $\phi_r \approx 1.4$. This demonstrates that the gauge transformation provides an effective means of tuning the non-normality of the HEOM Liouvillian, thus offering a path to improving the robustness and stability of long-time HEOM simulations. 

The gauge also offers control over the spectral properties of the HEOM Liouvillian.
Specifically, Fig.~\ref{fig1}(d) shows the maximum real part of the eigenvalue spectrum of $\bf A$ as a function of $\phi_r$. As $\phi_r$ is varied, the largest real eigenvalue can be shifted from the positive to the negative half-plane. This spectral shift suppresses numerical divergence and improves the stability of HEOM propagation. In this parameter regime, both the Henrici departure from normality and the maximum real part of the eigenvalue spectrum are significantly reduced at $\phi_r \approx 1.4$. This indicates that the gauge transformation simultaneously suppresses non-normality and shifts the spectrum toward the stable half-plane, improving numerical stability. Figure~\ref{fig1}(d) compares the eigenspectra for $\phi_r = 1.0$ and $\phi_r = 1.4$. For $\phi_r = 1.0$, several eigenvalues possess positive real parts, whereas for $\phi_r = 1.4$ the spectrum becomes more compact and shifts toward the stable half-plane. The reduced spectral spread suggests a reduction in stiffness, while the absence of eigenvalues with positive real contributions improves numerical stability. 

\begin{figure}[b]
\centering
\includegraphics[width=5.00cm]{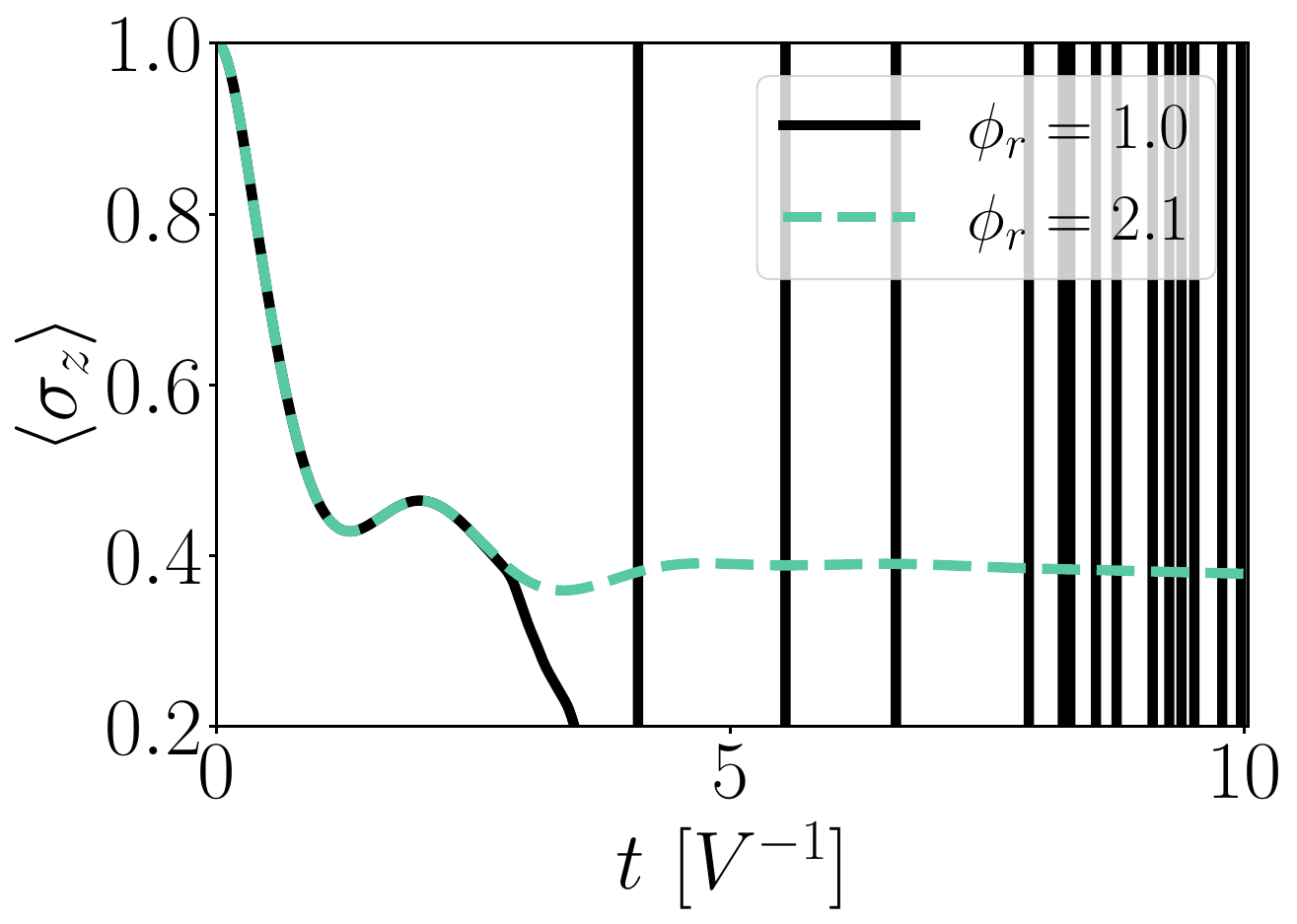}
\vspace{-10pt}
\caption{
$\langle\sigma_z(t)\rangle$ with strong coupling to a BO ($p_0=10.0$) obtained with conventional complex-pole HEOM ($\phi_r =1.0$) and gauge-transformed HEOM ($\phi_r =2.1$).
}
\vspace{-10pt}
\label{fig2}
\end{figure}

The gauge that minimizes the Henrici departure from normality and ensures that all real parts of the Liouvillian eigenspectrum remain negative defines our GO-HEOM. In practice, the optimal gauge is obtained by scanning the gauge parameter $\phi_r$ and selecting the value that yields the most stable numerical propagation.

Having established the validity and protocol for identifying the case-dependent GO-HEOM, we can now interrogate its benefits in a traditionally challenging parameter regime of strong coupling ($p_0=10.0$). Here, the conventional complex-pole HEOM ($\phi_r=1.0$) exhibits eigenvalues with positive real parts. Consequently, conventional HEOM becomes numerically unstable, eventually diverging (see Fig.~\ref{fig2}). In our GO-HEOM ($\phi_r=2.1$), all unstable modes disappear and the dynamics remains stable throughout the simulation, substantially extending the range of stable HEOM calculations.

The sub-Ohmic spin-boson model exhibits paradigmatic quantum criticality spanning delocalized and localized phases at weak and strong coupling, respectively~\cite{winter2009quantum}, whose dynamical signatures have proven difficult to access with numerically exact methods at the coupling strengths where the transition occurs. We show that GO-HEOM enables stable, exact dynamics precisely in this regime. Sub-Ohmic spectral densities, given by $J(\omega)=\frac{\pi}{2}\alpha \omega_c^{1-s}\omega^s e^{-\omega/\omega_c}$~\cite{leggett87,bulla03,duan17,wang10a,goulko2025transient,gong2026entanglement}, are numerically challenging because their low-frequency-dominated structure generates many slowly decaying bath memory modes, leading to long-time non-Markovian correlations and significantly increasing the required hierarchy depth~\cite{hartmann2017exact,xu22}.
While complex-pole HEOM offers more stable and less noisy long-time dynamics than TD-DMRG and ML-MCTDH for $s=0.5$~\cite{xu22}, it fails for more strongly sub-Ohmic regimes. We target this difficult regime by setting $s=0.3$, $k_B T = 0$, $\omega_c = 20$, while varying the coupling strength from $\alpha=0.02$ to $\alpha=0.2$. This strongly sub-Ohmic exponent places the system deep in the challenging long-memory regime, providing a challenging simulation for HEOM methods. See SM for our optimized $\phi_r$ for each coupling strength.

Figure~\ref{fig3}(a) demonstrates that our GO-HEOM enables stable simulations over a wide range of coupling strengths, from the weak-coupling regime to the strongly coupled regime ($\alpha = 0.2$). The resulting dynamics clearly captures the quantum phase transition from the delocalized to the localized regime as the coupling strength increases~\cite{bulla03,winter2009quantum}. In fact, GO-HEOM enables convergent simulations up to and beyond $ \alpha = 0.2$, extending the accessible coupling range by a factor of 4 beyond what has been previously achievable with complex-pole HEOM methods~\cite{xu22}. Figure~\ref{fig3}(b) compares the dynamics obtained from the conventional complex-pole HEOM ($\phi_r = 1.0$) and our GO-HEOM in the strong coupling regime ($\alpha = 0.2$). In contrast to the BO spectral density discussed above, the numerical difficulties in the sub-Ohmic case are not primarily associated with eigenvalues in the unstable half-plane. Instead, the long-lived memory modes of the sub-Ohmic environment significantly enhance the non-normality of the HEOM Liouvillian, leading to transient amplification and accumulation of numerical errors. As a result, the conventional formulation progressively deviates from the stable transformed dynamics at long times.

\begin{figure}[t]
\centering
\includegraphics[width=8.70cm]{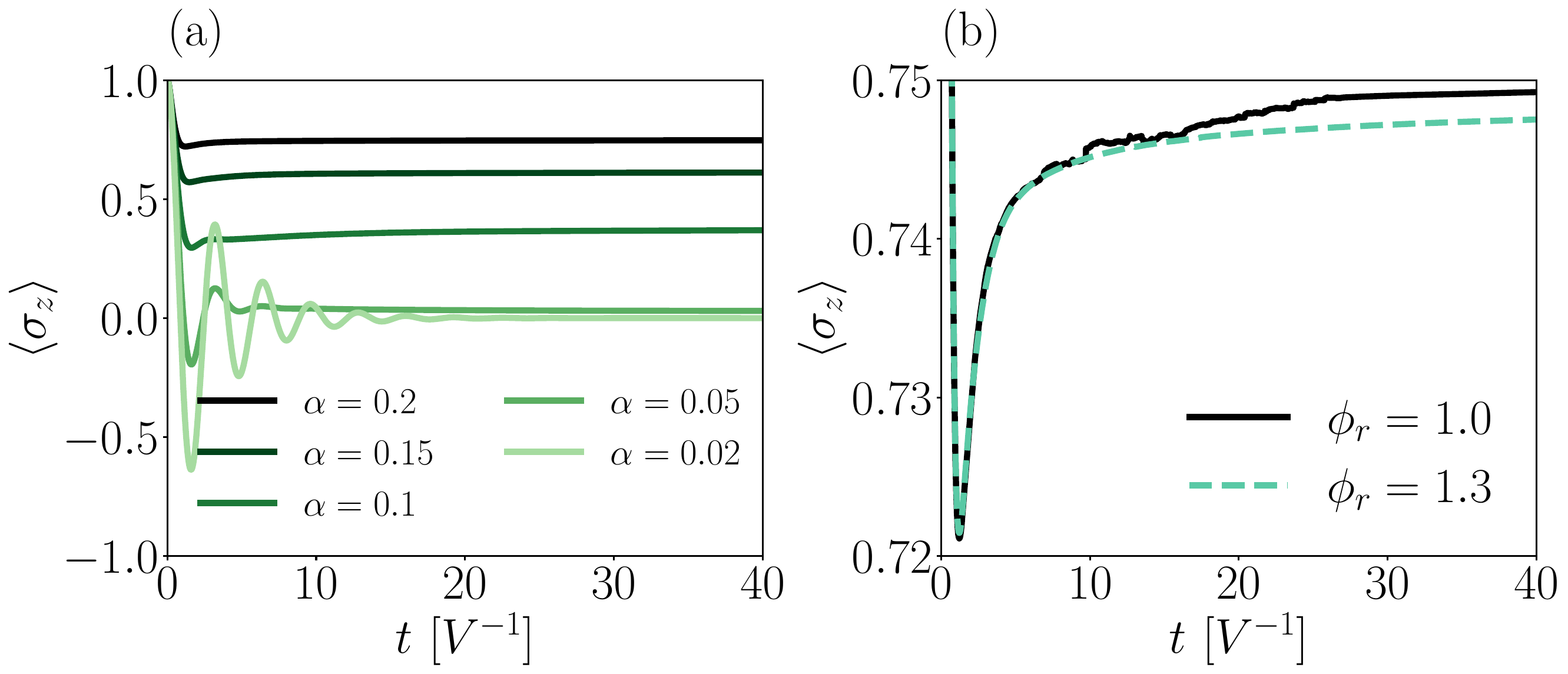}
\vspace{-20pt}
\caption{ Spin-boson dynamics with a sub-Ohmic spectral density at zero temperature. (a) $\langle\sigma_z(t)\rangle$ as a function of coupling strength, $\alpha$. (b) Error accretion in conventional HEOM ($\phi_r=1.0$) compared to our GO-HEOM ($\phi_r=1.3$) under strong coupling ($\alpha=0.2$).
}
\vspace{-10pt}
\label{fig3}
\end{figure}
\textit{Conclusion} --- Our work reveals that complex-pole HEOM possesses a previously unknown gauge freedom that preserves the bath correlation function while continuously reshaping the structure of the HEOM Liouvillian. By exploiting this freedom, we established a direct link between the gauge, eigenspectrum, and non-normality of the hierarchy generator, and its resulting numerical stability. This gauge offers a simple but effective mechanism for controlling both spectral instability and non-normal amplification. For the BO spectral density, the transformation shifts unstable eigenvalues into the stable half-plane and eliminates numerical divergence in the strong-coupling regime. For the sub-Ohmic spectral density, it suppresses non-normality-induced error amplification arising from long-lived bath memory modes, substantially improving long-time accuracy. These results reveal that spectral instability and non-normal amplification constitute two distinct mechanisms limiting the stability of complex-pole HEOM.

More broadly, our work identifies the structure of the HEOM Liouvillian as an important degree of freedom for improving numerical performance. Because the gauge transformation is independent of the particular decomposition scheme used to obtain the bath correlation function, it can be combined with existing complex-pole approaches, including Padé, Prony, and AAA decompositions, and additional efficiency-enhancement strategies. We therefore expect GO-HEOM to provide a general route toward stable simulations of strongly coupled and highly non-Markovian open quantum systems, directly compatible with existing efficiency enhancements. More broadly, our identification of gauge freedom in HEOM raises the question of whether analogous representational redundancies exist in other exact methods---influence functional approaches~\cite{makarov94,makarov95,fux23}, the hierarchy of pure states~\cite{suess14,hartman17} and tensor-network propagators among them---and whether they too can be exploited for numerical gain. The structure of the Liouvillian, long treated as fixed by physics, may prove to be a general target for optimization.

\section*{acknowledgement}

A.M.C.~and T.L.~were supported by an Early Career Award in the CPIMS program in the Chemical Sciences, Geosciences, and Biosciences Division of the Office of Basic Energy Sciences of the U.S.~Department of Energy under Award DE-SC0024154. A.M.C.~also acknowledges the support from a David and Lucile Packard Fellowship for Science and Engineering. We thank Anthony J Dominic III, Matthew Laskowski, Pranay Venkatesh, Prof.~Hsing-ta Chen and Prof.~Qiang Shi for discussion and comments on the manuscript. This work utilized the Alpine high-performance computing resource at the University of Colorado Boulder. Alpine is jointly funded by the University of Colorado Boulder, the University of Colorado Anschutz, Colorado State University, and the National Science Foundation (Award No. 2201538).

\bibliography{quantum}

\pagebreak

\end{document}


\title{Supplemental Material for\\``Hidden Gauge Freedom in\\ Complex-Pole Hierarchical Equations of Motion''}

\author{Tianchu Li}
\affiliation{Department of Chemistry, University of Colorado Boulder, Boulder, Colorado 80309, USA\looseness=-1}
\author{Andr\'es Montoya-Castillo}
\email{Andres.MontoyaCastillo@colorado.edu}
\affiliation{Department of Chemistry, University of Colorado Boulder, Boulder, Colorado 80309, USA\looseness=-1}
\maketitle

\pagebreak
\section{Computational details for HEOM}

To improve computational efficiency, we employ a matrix product state (MPS) representation of HEOM~\cite{shi18,ke22,borrelli21}.
Within this framework, the hierarchy is truncated by imposing a maximum occupation number $N_b$ for each bath mode, such that auxiliary density operators (ADOs) with $n_k \ge N_b$ are discarded.

The ADOs are represented as
\begin{align}
    \rho_{\mathbf n}
    =
    A_s B^{[0]} B^{[1]} \cdots B^{[K-1]},
\end{align}
where $A_s$ denotes the system tensor obtained through the Choi transformation and $B^{[i]}$ is the tensor associated with the $i$th bath mode.

The resulting MPS representation allows the HEOM dynamics to be propagated efficiently using the time-dependent variational principle (TDVP)~\cite{lubich15}.
Within the TDVP framework, the exact HEOM equations of motion are projected onto the tangent space of the variational MPS manifold, enabling bond--fixing time evolution. To further enhance computational performance, all tensor contractions and linear-algebra operations are accelerated using GPUs.

For all simulations, the initial state is chosen as the product state
\begin{align}
    \rho(0)
    =
    |0\rangle\langle 0|
    \otimes
    \frac{e^{-\beta H_{\rm B}}}
         {Z},
\end{align}
corresponding to the system prepared in the excited state and the bath in thermal equilibrium.
During the time evolution, we measure the population difference
\begin{align}
    \langle \sigma_z(t)\rangle
    =
    {\rm Tr}
    \big[
        \sigma_z \rho(t)
    \big].
\end{align}

We employ the adaptive Antoulas--Anderson (AAA) algorithm~\cite{xu22} to obtain an exponential decomposition of the bath correlation function.
The AAA algorithm constructs a rational approximation to the noise spectrum
\begin{align}
    S(\omega)
    =
    \int_{-\infty}^{\infty}
    dt\, e^{i\omega t} C(t),
\end{align}
where $C(t)$ is the bath correlation function.
Its central idea is to represent $S(\omega)$ with a minimal number of poles while maintaining a prescribed accuracy.
This is achieved through the barycentric representation
\begin{align}
    \tilde S(\omega)
    =
    \sum_{j=1}^{m}
    \frac{W_j S(\Omega_j)}
         {\omega-\Omega_j}
    \Big/
    \sum_{j=1}^{m}
    \frac{W_j}
         {\omega-\Omega_j},
\end{align}
which satisfies the interpolation condition
$\tilde S(\Omega_j)=S(\Omega_j)$.
The fitting procedure is terminated when the relative approximation error falls below a prescribed tolerance $\varepsilon$.
Further details can be found in Ref.~\onlinecite{xu22}.

For the Brownian oscillator (BO) spectral density, we use a fitting tolerance of $\varepsilon=10^{-6}$, yielding an exponential decomposition with three terms.
For the sub-Ohmic spectral density, the same tolerance results in twenty-five exponential terms for all coupling strengths considered in this work.

\begin{table}[h]
\centering
\caption{Hierarchy truncation parameter $N_b$ and gauge-scaling parameter $\phi_r$ used for different coupling strengths $\alpha$.}
\begin{tabular}{c c c c c c}
\hline
$\alpha$ & 0.02 & 0.05 & 0.10 & 0.15 & 0.20 \\
\hline
$N_b$    & 10   & 15   & 20   & 20   & 25   \\
$\phi_r$ & 1.1  & 1.5  & 1.3  & 1.3  & 1.3  \\
\hline
\end{tabular}
\label{tab:subohmic_parameters}
\end{table}

\begin{figure}[htbp]
\renewcommand{\figurename}{FIG.}              
\renewcommand{\thefigure}{S\arabic{figure}}
\centering
\includegraphics[width=15cm]{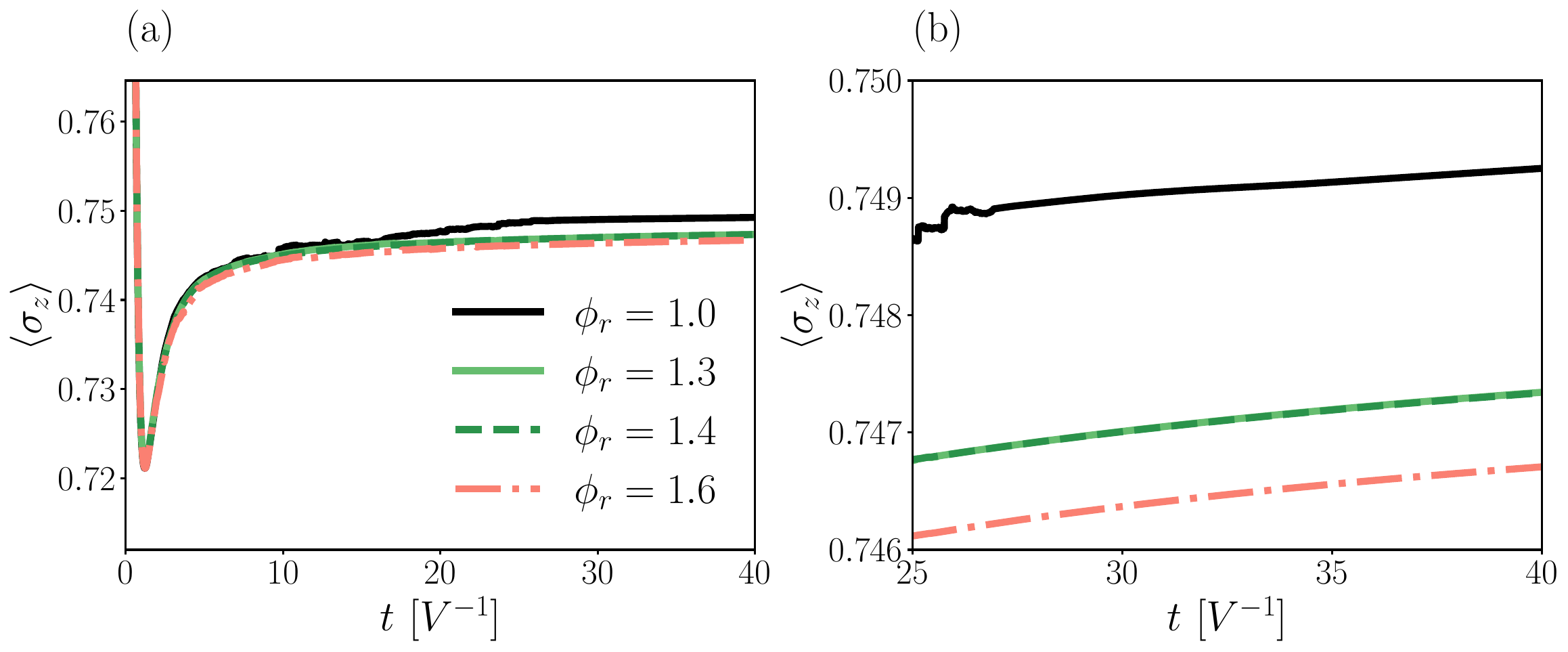}
\vspace{1em}
\caption{(a) Spin-boson dynamics with a sub-Ohmic spectral density at zero temperature in strong coupling regime ($\alpha=0.2$) with different gauge-scaling parameter. (b) Magnified view of the long-time dynamics in panel (a).
}
\label{figs1}
\end{figure}

\begin{figure}[htbp]
\renewcommand{\figurename}{FIG.}              
\renewcommand{\thefigure}{S\arabic{figure}}
\centering
\includegraphics[width=15cm]{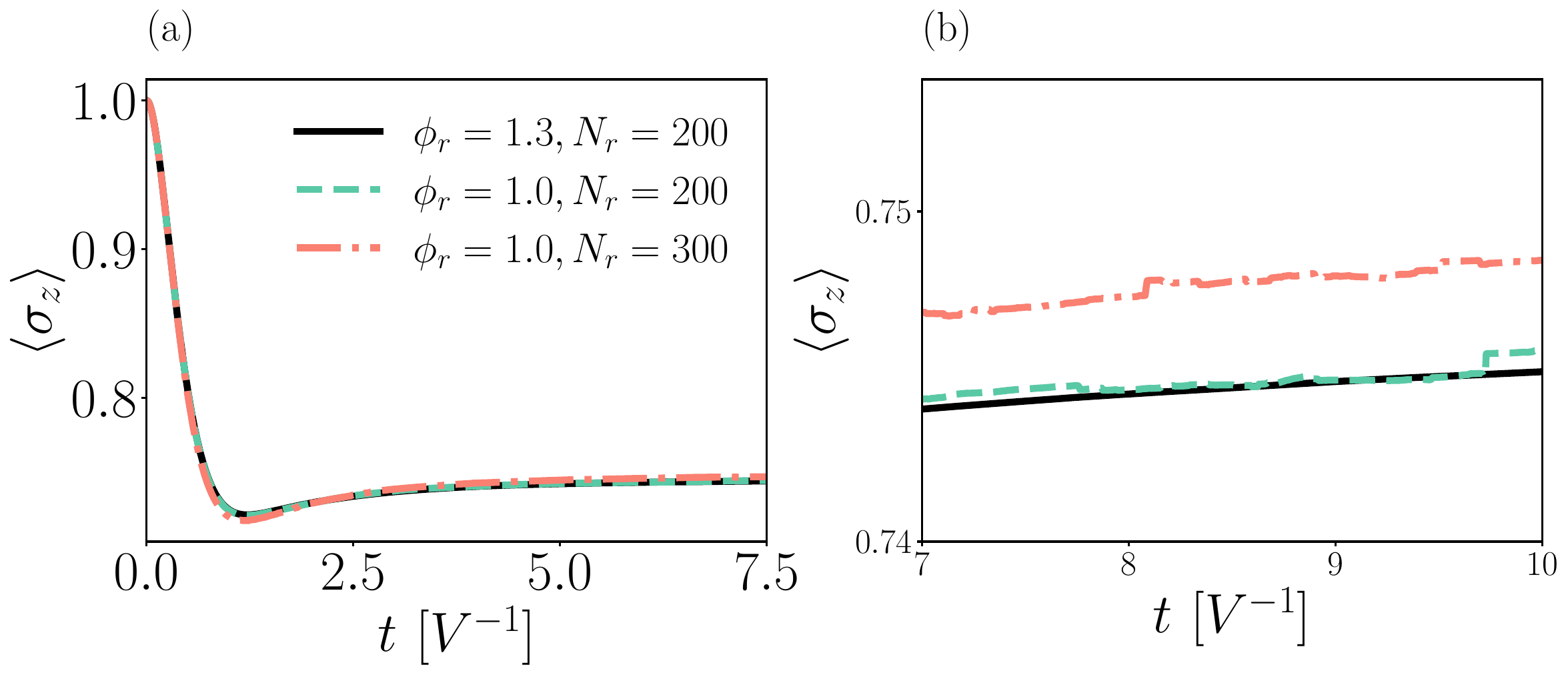}
\vspace{1em}
\caption{(a)Spin-boson dynamics with a sub-Ohmic spectral density at zero temperature in strong coupling regime ($\alpha=0.2$) with different gauge-scaling parameter and MPS bond dimension $N_r$. (b) Magnified view of the long-time dynamics in panel (a).
}
\label{figs2}
\end{figure}

Table~\ref{tab:subohmic_parameters} lists the hierarchy truncation parameter $N_b$ and the optimized gauge-scaling parameter $\phi_r$ for each coupling strength $\alpha$ in sub--Ohmic spectral density. Figure~\ref{figs1} compares the dynamics obtained with different values of $\phi_r$ in the strong-coupling regime ($\alpha=0.2$). The dynamics corresponding to $\phi_r=1.3$ and $1.4$ coincide, demonstrating convergence with respect to the gauge parameter. By contrast, values of $\phi_r$ away from this optimal region lead to stronger error amplification and ultimately unreliable long-time dynamics. Figure~\ref{figs2} shows that increasing the MPS bond dimension substantially reduces the TDVP propagation error but does not eliminate the sawtooth error, indicating that the latter does not originate from the tensor-network truncation.


\begin{figure}[htbp]
\renewcommand{\figurename}{FIG.}              
\renewcommand{\thefigure}{S\arabic{figure}}
\centering
\includegraphics[width=8cm]{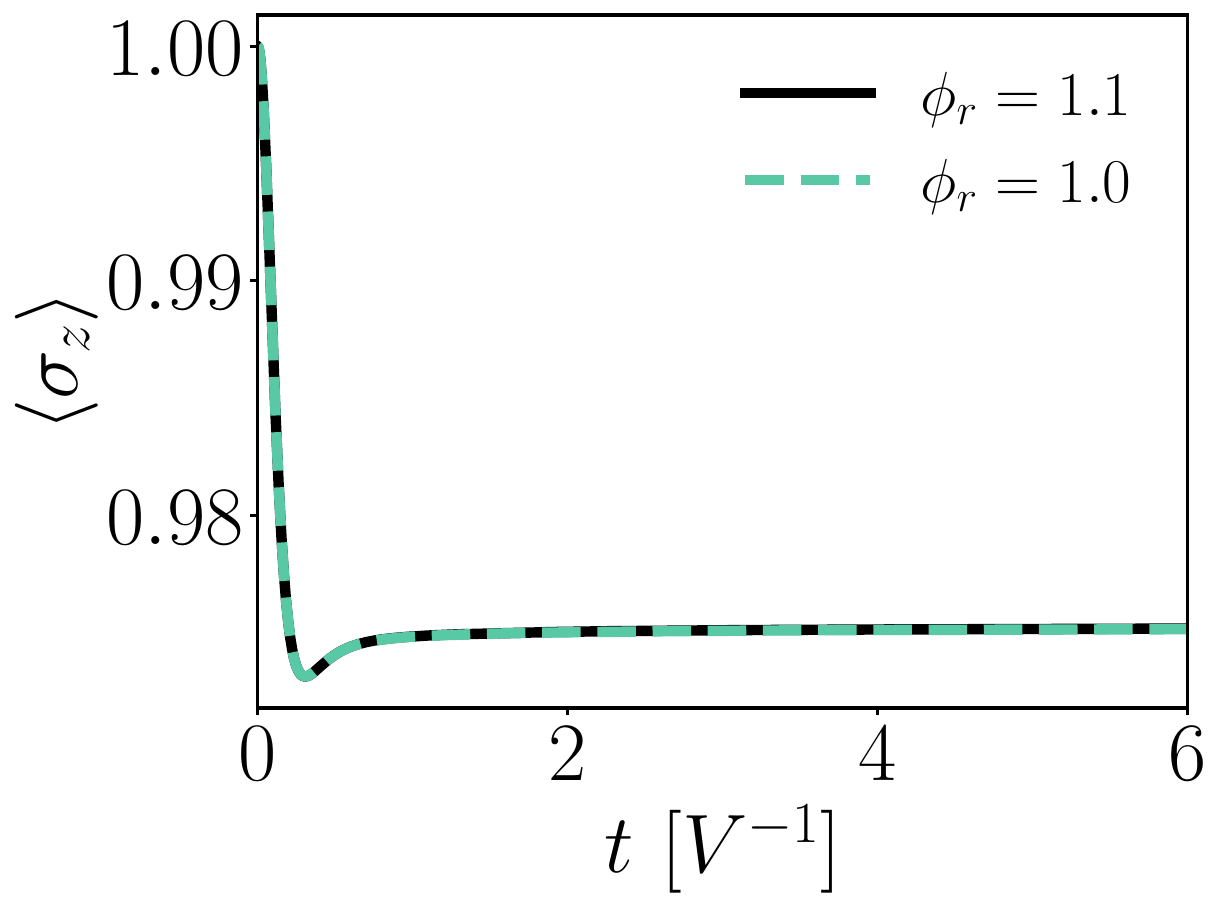}
\vspace{1em}
\caption{Spin-boson dynamics with a sub-Ohmic spectral density ($s=0.5$) at zero temperature in the strong-coupling regime ($\alpha=1.0$), comparing the conventional complex-pole HEOM and GO-HEOM. 
}
\label{figs3}
\end{figure}

Finally, as a benchmark against previous complex-pole HEOM results, we consider the sub-Ohmic case with $s=0.5$ at the strongest coupling strength($\alpha=1.0$) previously investigated using the conventional complex-pole HEOM. In that parameter regime, the conventional complex-pole HEOM was previously benchmarked against numerically exact methods, including, Time-Dependent Density Matrix Renormalization Group(TD-DMRG), and Multi-Layer Multiconfiguration Time-Dependent Hartree(ML-MCTDH), showing excellent agreement~\cite{xu22}. Figure~\ref{figs3} compares the dynamics obtained from the conventional complex-pole HEOM and GO-HEOM. The two approaches produce indistinguishable results, confirming that GO-HEOM faithfully reproduces the established benchmark dynamics while providing a more robust numerical representation.

\section{Phase-space representation of the complex-pole HEOM}

Here we provide the explicit form of the
complex-pole HEOM defined in the previous work~\cite{xu22},
\begin{align}
     \dot{\rho}_{\mathbf m, \mathbf n}&=-\left(i\mathcal L+\sum_k z_k m_k+\sum_k z_k^* n_k\right)\rho_{\mathbf m,\mathbf n}-i\sum_k\left[\sigma_z,\rho_{\mathbf m_k^+,\mathbf n}\right]\nonumber\\
     &-i\sum_k\left[\sigma_z,\rho_{\mathbf m,\mathbf n_k^+}\right]-i\sum_k d_k m_k\,\sigma_z\rho_{\mathbf m_k^-,\mathbf n}+i\sum_k d_k^* n_k\,\rho_{\mathbf m,\mathbf n_k^-}\sigma_z,
\label{eq:heom0}
\end{align}
where $\mathbf m={m_1,m_2,\dots}$ and $\mathbf n={n_1,n_2,\dots}$ are integer indices labeling the ADOs. 

The HEOM was previously derived by expanding the Wigner distribution function associated with a phase-space equation in a harmonic-oscillator basis~\cite{shi09c,liu14,yan20,li22}. We adopt the same representation but in reverse: starting from the complex-pole HEOM, we introduce a phase-space representation by expressing the Wigner distribution function in the same harmonic-oscillator basis used in the original derivation of the HEOM,
\begin{align}
    \rho_s(x,p;t)=\sum_{m,n}\frac{1}{\sqrt{m!n!}}\rho_{m,n}(t)\phi_{m,n}(x,p),
\end{align}
where
\begin{align}
    \phi_{m,n}(x,p)=\frac{1}{\sqrt{2\pi}}e^{-S}(c_1^+)^m(c_2^+)^n|0,0\rangle,
\end{align}
with $S = x^2/4+p^2/4$ and $|0,0\rangle = e^{-S}$. Since only the vacuum basis function has a nonzero phase-space integral,
\begin{align}
    \rho_s(t)=\int dxdp\,\rho_s(x,p;t)=\rho_{0,0}(t),
\end{align}
showing that $\rho_{0,0}$ corresponds to the physical reduced density matrix.

With the above definition, the hierarchy indices can be represented as differential operators acting on the phase-space distribution. In particular, the diagonal, lowering, and raising terms in the HEOM hierarchy are mapped as follows:
\begin{align}
    m\rho_{m,n} &\mapsto e^{-S}(c_1^+c_1)e^S\rho_s(x,p),\nonumber \\
    \rho_{m-1,n} &\mapsto e^{-S}(c_1^+)e^S\rho_s(x,p),\nonumber \\
    \rho_{m+1,n} &\mapsto e^{-S}(c_1)e^S\rho_s(x,p).
\end{align}
Because the ladder operators act on the basis functions rather than on the expansion coefficients ($\rho_{m,n}$), raising (lowering) a basis index corresponds to lowering (raising) the associated hierarchy index. Analogous relations hold for the second hierarchy index $n$ through $c_2$ and $c_2^+$.

Using the mappings above, we recast the complex-pole HEOM \textit{exactly} into the equivalent phase-space equation of motion,
\begin{align}
    \dot{\rho}_s(x,p) =&\;-e^{-S}                 \Bigl(i\mathcal{L} + z_0{c}_{1}^+{c}_{1} + z_0^*{c}_{2}^+{c}_{2}\Bigr) e^{S}\rho_s(x,p) -ie^{-S}\bigl({c}_{1}+{c}_{2}\bigr)e^{S} \bigl[\sigma_z,\rho_s(x,p)\bigr] \nonumber\\ &-ie^{-S} \Bigl[ d_0\,{c}_{1}^+ e^{S} \sigma_z\,\rho_s(x,p) -d_0^*{c}_{2}^+ e^{S} \rho_s(x,p)\,\sigma_z \Bigr]. 
    \label{eq:pseom} 
\end{align} 
To reveal the underlying Fokker--Planck structure, we apply the linear transformation defined in Eqs.~(6a) and (6b) of the main text, which maps the complex ladder operators $(c_1,c_1^+,c_2,c_2^+)$ onto a set of real bosonic creation and annihilation operators $(a,a^+,b,b^+)$ satisfying
\begin{subequations} 
\begin{align} 
    \hat a&=\frac{\partial}{\partial x}+\frac{x}{2},\quad&\hat b&=\frac{\partial}{\partial p}+\frac{p}{2},\\ \hat a^+&=-\frac{\partial}{\partial x}+\frac{x}{2},\quad&\hat b^+&=-\frac{\partial}{\partial p}+\frac{p}{2}. 
\end{align} 
\end{subequations} 
Under the standard inner product, where operators act to the right,
\begin{align}
\langle f|\hat{O}|g\rangle
=
\int dx\,dp\, f^*(x,p)O(x, p, \partial_x, \partial_p)g(x,p),
\end{align}
integration by parts yields
\begin{align}
    \left(\frac{\partial}{\partial x}\right)^+
    =
    -\frac{\partial}{\partial x},
    \qquad
    \left(\frac{\partial}{\partial p}\right)^+
    =
    -\frac{\partial}{\partial p}.
\end{align}
Therefore, $a^+$ and $b^+$ are the Hermitian conjugates of $a$ and $b$, respectively. Substituting these operators into Eq.~(\ref{eq:pseom}) yields 
\begin{align} 
    \dot{\rho}_s(x,p) =&-e^{-S} \Bigl(i\mathcal{L} + \mathcal L_{\rm FP}\Bigr) e^{S}\rho_s(x,p) -ie^{-S} b e^{S} \bigl[\sigma_z,\rho_s(x,p)\bigr] \nonumber\\ &-ie^{-S} \left[ d_0\left(b^+-\frac{\gamma-i\omega}{\Omega}a^+\right) e^{S} \sigma_z\,\rho_s(x,p) -d_0^*\left(b^+-\frac{\gamma+i\omega}{\Omega}a^+\right) e^{S} \rho_s(x,p)\,\sigma_z \right], 
    \label{eq:newphase}
\end{align}
where
\begin{align}
    \tilde{\mathcal L}_{\rm FP}&=-e^{-S}\Bigl(2\gamma b^+b+\Omega ab^+-\Omega a^+b\Bigr)e^{S}\nonumber=
    2\gamma \left(p\frac{\partial}{\partial p}+\frac{\partial^2}{\partial p^2}\right)-\Omega\left(p\frac{\partial}{\partial x}+x\frac{\partial}{\partial p}\right).
\end{align}
This is precisely the Fokker--Planck operator that appears in the quantum Fokker--Planck equation (QFPE)~\cite{garg85,li22}. The QFPE describes the reduced dynamics of a system linearly coupled to a collective harmonic mode, which linearly couples to an external bath with Ohmic spectral density
\begin{align}
    J(\omega)=2\gamma\omega .
\end{align}
Within this mapping, the phase-space variables $x$ and $p$ correspond to the coordinate and momentum of the collective harmonic mode that directly couples to the system. Consequently, Eq.~(\ref{eq:newphase}) may be viewed as a generalized QFPE. While the conventional QFPE corresponds to a collective mode coupled to an Ohmic bath, the phase-space equation of motion in Eq.~(\ref{eq:newphase}) describes an analogous collective mode interacting with an environment characterized by a general spectral density. In this sense, the complex-pole HEOM provides a systematic extension of the QFPE framework beyond the Ohmic limit.

\section{Coefficients in Equation~(9)}

The coefficients appearing in Eq.~(10) of the main manuscript are defined as
\begin{align}
    A_{11}&=\frac{i}{\sin\phi}\left(\Omega-\gamma e^{i\phi}\right),\\
    A_{22}&=\frac{i}{\sin\phi}\left(\gamma e^{-i\phi}-\Omega\right),\\
    A_{12}&=\frac{i e^{i\phi}}{\sin\phi}\left(\Omega\cos\phi-\gamma\right),\\
    A_{21}&=-\frac{i e^{-i\phi}}{\sin\phi}\left(\Omega\cos\phi-\gamma\right),
\end{align}
where
\begin{align}
\Omega=\sqrt{\gamma^2+\omega^2}.
\end{align}
For $\phi=\theta$, where
\begin{align}
\cos\theta=\frac{\gamma}{\Omega},
\qquad
\sin\theta=\frac{\omega}{\Omega},
\end{align}
the off-diagonal coefficients vanish,
\begin{align}
A_{12}=A_{21}=0,
\end{align}
while
\begin{align}
A_{11}=z=\gamma+i\omega,
\qquad
A_{22}=z^*=\gamma-i\omega,
\end{align}
thus recovering the conventional complex-pole HEOM.

\section{Invariance under Sign Reversal of the Gauge Parameters}

Consider a sign reversal of a single gauge parameter,
\begin{align}
\phi_k \rightarrow -\phi_k,
\end{align}
while keeping all other gauge parameters unchanged. Under this transformation, the ADOs satisfy the relation
\begin{align}   
    \tilde\rho_{\mathbf m,\mathbf n}=(-1)^{m_k+n_k}\rho_{\left(m_0,\ldots,n_k,\ldots,m_K\right);\left(n_0,\ldots,m_k,\ldots,n_K\right)},
\end{align}
where the indices $m_k$ and $n_k$ associated with the transformed mode are exchanged. Hence, the HEOM are symmetric with respect to the sign of the gauge parameter, with this sign change translating into a simple exchange of ADO indices.

\section{Deriving gauge-transformed HEOM through a generalized bath correlation function decomposition}

To establish a direct connection between gauge-transformed HEOM and a bath-correlation-function decomposition, we introduce two time-dependent basis functions to decompose the bath correlation function, which satisfies
\begin{equation}
    \frac{d}{dt}
    \begin{pmatrix}
    \xi_1(t)\\
    \xi_2(t)
    \end{pmatrix}
    =
    -{\bf A}
    \begin{pmatrix}
    \xi_1(t)\\
    \xi_2(t)
    \end{pmatrix},
\end{equation}
where ${\bf A}$ is defined in Eq.~(10) of the main manuscript. For the initial condition $\xi_1(0)=\xi_2(0)=1$, one obtains

\begin{align}
    \xi_1(t)&=e^{-\gamma t}\left[\cos(\omega t)+i\frac{\Omega-\gamma\cos\phi+e^{i\phi}\left(\Omega\cos\phi-\gamma\right)}{\omega\sin\phi}\sin(\omega t)\right],\\
    \xi_2(t)&=e^{-\gamma t}\left[\cos(\omega t)-i\frac{\Omega-\gamma\cos\phi+e^{-i\phi}\left(\Omega\cos\phi-\gamma\right)}{\omega\sin\phi}\sin(\omega t)\right].
\end{align}

These basis functions are related to the conventional exponential modes through

\begin{equation}
\begin{pmatrix}
\xi_1(t)\\
\xi_2(t)
\end{pmatrix}
=
{\bf M}(\phi)
\begin{pmatrix}
\eta_1(t)\\
\eta_2(t)
\end{pmatrix},
\end{equation}

where

\begin{align}
\eta_1(t)&=e^{-z_0 t},
\\
\eta_2(t)&=e^{-z_0^{*} t}.
\end{align}

Using this relation, the single-exponential bath correlation function, $C(t)=d_0 e^{-z_0 t}$, can be equivalently expressed in terms of $\xi_1(t)$ and $\xi_2(t)$ as
\begin{align}
C(t)
=
-\frac{i d_0}{2\sin\phi}
\left[
\left(e^{i\phi}-e^{-i\theta}\right)\xi_1(t)
+
\left(-e^{-i\phi}+e^{-i\theta}\right)\xi_2(t)
\right].
\end{align}
This representation shows a generalized bath-correlation-function decomposition that is analytically equivalent to the conventional complex-pole expansion. Following the standard influence-functional derivation of HEOM~\cite{tanimura89,tanimura06}, we define the ADOs as
\begin{align}
    &\rho_{m,n}(t)=\mathcal D[q^+(t)]\,\mathcal D[q^-(t)]\,
    e^{i\left\{S_{+}[q^{+}(t)]-S_{-}[q^{-}(t)]\right\}}
    \nonumber\\
    &\times
    \prod_k
    \left\{
    -i\int_{0}^{t}d\tau
    \frac{d}{2\sin\phi}
    \left[
    q^{+}(\tau)
    \left(e^{i\phi}-e^{-i\theta}\right)
    \xi_{1,k}(t-\tau)
    -
    q^{-}(\tau)
    \left(e^{i\phi}-e^{i\theta}\right)
    \xi_{1,k}(t-\tau)
    \right]
    \right\}^{m_k}
    \nonumber\\
    &\times
    \prod_k
    \left\{
    -i\int_{0}^{t}d\tau
    \frac{d}{2\sin\phi}
    \left[
    q^{+}(\tau)
    \left(e^{-i\theta}-e^{-i\phi}\right)
    \xi_{2,k}(t-\tau)
    -
    q^{-}(\tau)
    \left(e^{i\theta}-e^{-i\phi}\right)
    \xi_{2,k}(t-\tau)
    \right]
    \right\}^{n_k}
    \nonumber\\
    &\times
    \mathcal F\!\left[q^{+}(t),q^{-}(t)\right]
    \rho(0).
\end{align}
Here, $\{ q^{\pm}(t)\}$ are the forward and backward path and $\mathcal F$ is the Feynman Vernon influence functional~\cite{feynman63}. Taking the time derivative of the ADOs and following the standard HEOM derivation yields Eq.~(11) of the main text. This demonstrates that gauge-transformed HEOM can be derived directly from a generalized bath correlation function decomposition without invoking the phase-space representation.

\section{General transformation in Equation~(7)}
Here, we consider the most general linear canonical transformation in Eq.~(8) with four degrees of freedom:
\begin{equation}
\begin{pmatrix}
\tilde c_1^+\\
\tilde c_2^+
\end{pmatrix}
=
\begin{pmatrix}
M_{11} & M_{12}\\
M_{21} & M_{22}
\end{pmatrix}
\begin{pmatrix}
a^+\\
b^+
\end{pmatrix},
\end{equation}
and
\begin{equation}
\begin{pmatrix}
\tilde c_1\\
\tilde c_2
\end{pmatrix}
=
\left(M^{-1}\right)^T
\begin{pmatrix}
a\\
b
\end{pmatrix}
=
\frac{1}{\Delta}
\begin{pmatrix}
M_{22} & -M_{21}\\
-M_{12} & M_{11}
\end{pmatrix}
\begin{pmatrix}
a\\
b
\end{pmatrix},
\end{equation}
where
\begin{equation}
\Delta=M_{11}M_{22}-M_{12}M_{21}.
\end{equation}
We require $\Delta\neq 0$, so that $M$ is invertible and the transformation is well defined. Under this condition, the canonical commutation relations are preserved, $[\tilde c_i,\tilde c_j^+]=\delta_{ij}$.

We can rewrite the Fokker--Planck operator as
\begin{align}
    2\gamma b^+b + \Omega(ab^+-a^+b)
    =
    A_{11}\tilde c_1^+ \tilde c_1
    +A_{22}\tilde c_2^+ \tilde c_2
    +A_{12}\tilde c_1^+ \tilde c_2
    +A_{21}\tilde c_2^+ \tilde c_1 ,
\end{align}
where the coefficients are defined as
\begin{align}
A_{11}
&=
\frac{
-2\gamma M_{21}M_{12}
-\Omega M_{21}M_{11}
-\Omega M_{22}M_{12}
}{\Delta},
\\
A_{22}
&=
\frac{
2\gamma M_{11}M_{22}
+\Omega M_{11}M_{21}
+\Omega M_{12}M_{22}
}{\Delta},
\\
A_{12}
&=
\frac{
-2\gamma M_{21}M_{22}
-\Omega M_{21}^2
-\Omega M_{22}^2
}{\Delta},
\\
A_{21}
&=
\frac{
2\gamma M_{11}M_{12}
+\Omega M_{11}^2
+\Omega M_{12}^2
}{\Delta}.
\end{align}
The corresponding HEOM can be written as:
\begin{align}
    \frac{d}{dt}\rho_{\mathbf m, \mathbf n} =& -\left(i\mathcal L + \sum_k A_{11,k} m_k +\sum_kA_{22,k}n_k \right)\rho_{\mathbf m, \mathbf n} -\sum_k \left(A_{12,k}m_k\rho_{{\bf m}_k^-,{\bf n}_k^+}+A_{21,k}n_k\rho_{{\bf m}_k^+,{\bf n}_k^-}\right)\nonumber \\
    &-i\sum_k d_k \left(\frac{M_{22,k}}{\Delta}e^{-i\phi}-\frac{M_{21}}{\Delta}\right)m_k\sigma_z \rho_{{\bf m}_k^-,\bf n} 
    +i \sum_k   d_k^*\left(\frac{M_{22,k}}{\Delta}e^{i\phi}-\frac{M_{21}}{\Delta}\right)m_k\rho_{{\bf m}_k^-,\bf n} \sigma_z\nonumber \\  
    &-i\sum_k d_k \left(\frac{M_{12,k}}{\Delta}e^{-i\phi}+\frac{M_{11}}{\Delta}\right)n_k\sigma_z \rho_{{\bf m},{\bf n}_k^-}
    +i \sum_k   d_k^*\left(\frac{M_{12,k}}{\Delta}e^{i\phi}+\frac{M_{11}}{\Delta}\right)n_k\rho_{{\bf m},{\bf n}_k^-} \sigma_z\nonumber \\
    &-i\sum_k M_{12,k}\left[\sigma_z, \rho_{{\bf m}_k^+,\bf n}\right]-i\sum_k M_{22,k}\left[\sigma_z, \rho_{{\bf m},{\bf n}_k^+}\right].
\end{align}
This fully general form contains four independent parameters for each complex pole. While consistent with the broader framework of equivalent HEOM representations discussed in Ref.~\cite{xu2023universal}, it provides a convenient parameterization of the gauge freedom in complex-pole HEOM and therefore defines a large gauge space. In practice, however, the resulting high-dimensional parameter space makes it difficult to systematically control or optimize the numerical behavior of the HEOM. For this reason, the main manuscript focuses on a restricted one-parameter subset of this transformation family.
\pagebreak
\bibliography{quantum}